\documentclass[12pt]{article}
\usepackage{latexsym}
\usepackage {graphicx}
\usepackage{cite}
\usepackage{amssymb}
\usepackage{epsfig}
\usepackage{hyperref}

\begin{document}

\begin{titlepage}
\title{Sliding with Friction and The Brachistochrone Problem}
\author{\large Alexander V. Kurilin{\footnote{E-mail address:
kurilin@mail.ru}}
\\ Moscow Technical University of Communications and Informatics (MTUCI)\\ 
\\}
\date{25 January 2025}
\maketitle 

\begin{abstract}
{We analyze the motion of a particle in the gravity field along a family of 
differentiable curves taking into account the Coulomb friction forces. A 
parametric equation of the optimal curve is given that generalizes the 
cycloid one in this case. The results of numerical calculations in the 
Mathcad program show that the found curve minimize the descent time for a 
given friction coefficient $\mu $ and can claim to be a brachistochrone with 
Coulomb friction.}
\end{abstract}

\end{titlepage}

\section{INTRODUCTION}

Johann Bernoulli's problem on the brachistochrone \cite{Courant,Tikhomirov}, formulated 1696, 
was a kind of challenge for his contemporaries and stimulated the 
development of new mathematics, which have found extensive applications in 
various fields of scientific knowledge. Almost all famous scientists of that 
time, including the author, were engaged in solving the problem of the 
brachistochrone. In the classical formulation, the essence of the 
brachistochrone problem is to find the fastest descent curve that connects 
some given points $A$ and $B$ lying in the vertical plane. It is assumed that the 
movement occurs without an initial speed under the influence of gravity 
along a smooth curve without friction. Distances $H$ and $L$ are given, that 
separate points $A$ and $B$ vertically and horizontally, respectively.

In recent years, the problem of the brachistochrone has once again attracted 
the attention of many researchers. We would especially like to mention the 
works when Coulomb friction forces were taken into account 
\cite{Ashby,Hayen,Sumbatov,Lipp,Barsuk}. In one of 
the first papers on this topic \cite{Ashby}, the authors discussed the fastest 
descent curve analytically and numerically employing variation principles. 
However, the equations obtained in that publications have implicit, rather 
complex form, which is extremely inconvenient for numerical analysis and 
mathematical modeling. In this regard, it would be interesting to calculate 
all the parameters of such movement using computer methods and trace how the 
speed of the particle, its coordinates and the total time of motion change 
depending on the shape of the trajectory and its curvature. Similar attempts 
were made in our works 
\cite{Kurilin2016,Kurilin-Arxiv,Kurilin2023,Kurilin2024}
for descent motion with Coulomb friction along 
parabolas and the cycloid. We calculated all the parameters of particle 
motion and realized that the cycloid does not provide the minimum descent 
time in such a case. It was also noted that the initial slope of the optimal 
slip curve graph couldn't be equal to that of a cycloid for all values of 
the friction coefficient $\mu $. Thus, the conclusion $\alpha _0 =\pi /2$, 
which has been employed in the papers \cite{Ashby,Sumbatov}, is groundless and erroneous 
from our point of view. This misconception arises from incorrect formulation 
of the variation problem, in which the boundary conditions imposed on the 
unknown function determining the shape of the optimal curve are included in 
the variation problem together with integral characteristics of the fastest 
descent curve. In this article it is proposed to separate the problem of 
finding the shape of the optimal descent curve from the task of calculating 
the optimal parameters of this curve that are determined from the boundary 
conditions. We present the results of such calculations for the optimal 
descent curve that claims to be a brachistochrone for a given value of the 
friction coefficient. We show that the statement $\alpha _0 (\mu )\le \pi /2$ 
provides the minimum descent time and compare various curves with Coulomb 
friction to confirm the name of brachistochrone.

\section{ANALYTICAL CALCULATIONS}

\begin{figure}[htbp]
\centerline{\includegraphics[width=130mm]{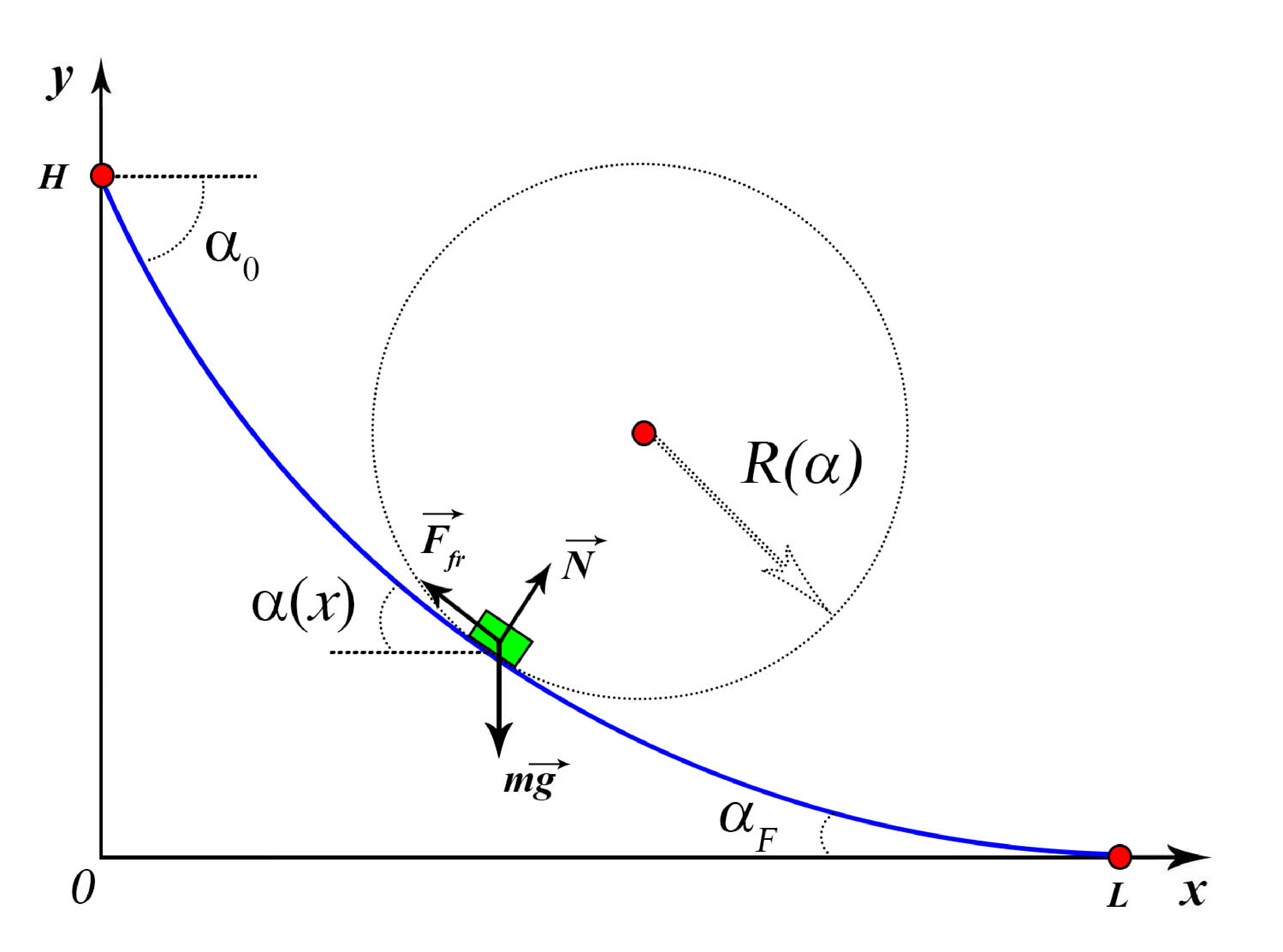}}
\caption[]{General view of the descent curve $y=\phi (x)$ in the 
gravity field when Coulomb friction force is taken into account.}
\label{Curve}
\end{figure}

Let's consider a small physical body, (a particle, material bead) of mass 
$m$ sliding down along some differentiable curve (thin rigid wire) specified 
by an explicit equation: $y=\phi (x)$. The classical laws of Newtonian 
mechanics make it possible to calculate the modulus of the sliding speed of 
the particle $\upsilon =\sqrt {\dot {x}^2+\dot {y}^2} $ at any point of the 
trajectory, if we take into account changes of potential energy in the 
gravity field and the work of friction forces over the entire previous 
section of motion. The elementary work of the Coulomb friction force through 
the infinite small displacement $ds$ can be expressed through the friction 
coefficient $\mu$; normal reaction $N$ and the friction force 
$F_{fr} =\mu N$, which in its turn depends on the curvature radius $R$ 
of the trajectory at a given point of view (see Fig. \ref{Curve}).

Explicitly, the modulus of the elementary work of the friction force can be 
written as follows:

\begin{equation}
\label{Dwork}
dA_{fr} =F_{fr} ds=\mu m\left( {g\cos \alpha +\frac{\upsilon ^2}{R}}
\right)ds.
\end{equation}
The inclination angle of the trajectory $\alpha \left( x 
\right)=-\mbox{arctan}\,\left[ {{\phi }'(x)} \right]$ at the observation 
point (see Fig. 1) is introduced in eq. (\ref{Dwork}). It can be calculated through 
the first derivative of the descent curve function: $y=\phi (x)$:
\begin{equation}
\label{SinCosA}
\sin\alpha\! \left( x \right)=\frac{-{\phi }'(x)}{\sqrt {1+{\phi
}'(x)^2} },\qquad \cos\alpha\! \left( x \right)=\frac{1}{\sqrt {1+{\phi
}'(x)^2} }.\,
\end{equation}
It will be convenient to express the sliding speed $\upsilon $ in terms of 
the horizontal projection of the velocity vector $\upsilon _x =\dot {x}$ 
using the relation:
\begin{equation}
\label{vel3}
\upsilon =\dot {x}\sqrt {1+{\phi }'(x)^2} =\frac{\dot {x}}{\cos \alpha}.
\end{equation}
Here, as usual, the prime at the function $y=\phi (x)$ indicates its 
derivative with respect to $x$-variable, when the dot indicates the 
derivative with respect to time. The curvature radius of the trajectory is 
expressed through the first and second derivatives of the sliding function 
$y=\phi (x)$:
\begin{equation}
\label{rad4}
\frac{1}{R(x)}=\frac{{\phi }''(x)}{\left[ {1+{\phi }'(x)^2}
\right]^{3/2}}={\phi }''(x)\cdot \cos ^3\alpha (x)
\end{equation}
Substituting formulas (\ref{vel3}), (\ref{rad4}) into (\ref{Dwork}), we obtain the expression for the 
friction force work through the infinitesimal displacement along the 
trajectory:
\begin{equation}
\label{Dwork5}
dA_{fr} =\mu m\left( {g+\upsilon ^2{\phi }''(x)\cos ^2\alpha } \right)\cos 
\alpha \cdot ds=\mu m\left( {g+\dot {x}^2{\phi }''(x)} \right)\cdot dx.
\end{equation}
The total work of resistance forces can be calculated by integrating formula 
(\ref{Dwork5}) over changes in horizontal $x$-coordinate from zero to the current 
location of the particle.
\begin{equation}
\label{work6}
A_{fr} =\mu m\int\limits_0^x {\left( {g+{\phi }''(x)\cdot W(x)}
\right)\cdot dx}
\end{equation}
In the resulting formula, an auxiliary function $W(x)$ has been introduced 
which is equal to the square of the horizontal projection of the velocity 
vector, expressed through the current $x$-coordinate:
\begin{equation}
\label{Wfun}
W(x)=\dot {x}^2(t)=\left( {\frac{dx}{dt}} \right)^2.
\end{equation}
Using the law of changes in the kinetic energy of a sliding particle along 
the chosen trajectory and taking into account the work of the friction 
forces (\ref{work6}), we arrive at the following expression:
\begin{equation}
\label{Energy8}
\frac{m\upsilon ^2}{2}=mg\left( {H-\phi (x)} \right)-\mu m\int\limits_0^x
{\left( {g+{\phi }''(x)\cdot W(x)} \right)\cdot dx}
\end{equation}
Reducing both sides of formula (\ref{Energy8}) by mass $m$ and passing with the help of 
(\ref{SinCosA}) and (\ref{vel3}) to the $x$-speed square function (\ref{Wfun}), we transform (\ref{Energy8}) into the 
following integral equation:
\begin{equation}
\label{IntegEq9}
\frac{1}{2}W(x)\left[ {1+{\phi }'(x)^2} \right]=g\left[ {H-\phi (x)}
\right]-\mu \int\limits_0^x {\left( {g+{\phi }''(x)\cdot W(x)}
\right)\cdot dx} .
\end{equation}
Differentiating both sides of formula (\ref{IntegEq9}) with respect to the $x$-variable, 
we obtain the first-order linear differential equation for the $W(x)$ 
function:
\begin{equation}
\label{Weqn10}
\frac{1}{2}\frac{dW(x)}{dx}=-\,\,\frac{{\phi }'(x)+\mu }{1+{\phi
}'(x)^2}\left[ {g+{\phi }''(x)W(x)} \right].
\end{equation}
Formula (\ref{Weqn10}) was obtained in our work \cite{Kurilin-Arxiv} directly from the Newton's laws 
of dynamics, where a general solution to this equation in quadratures was 
also found for an arbitrary twice differentiable descent curve $y=\phi (x)$:
\begin{equation}
\label{Wsol11}
W(x)=\frac{K(x)}{1+{\phi }'(x)^2}\exp \left[ {-2\mu \cdot 
\mbox{arctan}\left( {{\phi }'(x)} \right)} \right].
\end{equation}
The ``acceleration history'' function $K(x)$ satisfying the initial 
conditions $x(0)=0,\,\,\dot {x}(0)=0$ is introduced in equation (\ref{Wsol11}). The 
initial conditions imply that $K(0)=0$, so the function $K(x)$ can be written 
through the following integral:
\begin{equation}
\label{Kint12}
K(x)=-2g\int\limits_0^x {\left[ {\mu +{\phi }'(\xi )} \right]} \cdot \exp
\left[ {2\mu\, \mbox{arctan}\left( {{\phi }'(\xi )} \right)}
\right]d\xi
\end{equation}
Then the total descent time along a given trajectory $y=\phi (x)$ can be 
found by integrating the function (\ref{Wsol11}) along the horizontal coordinate.
\begin{equation}
\label{Time13}
T_0 =\int\limits_0^L {\frac{dx}{\dot {x}(t)}} =\int\limits_0^L 
{\frac{dx}{\sqrt {W\left( x \right)} }} .
\end{equation}
Formulas (\ref{Wsol11}) - (\ref{Time13}) were used in our papers 
\cite{Kurilin-Arxiv,Kurilin2024} to calculate the 
sliding time along parabolas and the cycloid taking into account the Coulomb 
friction forces. In this article we make an attempt to find the equation of 
the fastest descent curve (brachistochrone) for a given value of the 
friction coefficient $\mu $. Let us seek the equation of brachistochrone in 
the following parametric form:
\begin{equation}
\label{XY-alpha}
x(\alpha )=C \int\limits_{\alpha _0 }^\alpha {\psi (\beta )} \,d\beta
,\qquad y(\alpha )=H-C \int\limits_{\alpha _0
}^\alpha {\psi (\beta )\cdot \mbox{tan}\beta } \,\,d\beta ,
\end{equation}
where $\psi (\beta )$ is some unknown dimensionless function that determines 
the shape of the brachistochrone. Parameter $C$ is a scale factor having the 
dimension of length, $\alpha _0 $ is another free parameter corresponding to 
the initial inclination angle for the slip curve at the point $x=0\,$ 
$\Rightarrow  \quad \,\alpha _0 =\alpha \left( 0 
\right)=-\mbox{arctan}\,\,\left[ {{\phi }'(0)} \right]$. Equations (\ref{XY-alpha}) 
define the brachistochrone in parametric form, as a function of the angle 
parameter $\alpha $ which corresponds to the inclination of the curve graph 
$y=\phi (x)$ with respect to the $x$-axis (see Fig. \ref{Curve}). It is assumed that 
the brachistochrone is a curve with positive curvature (\ref{rad4}), for which the 
slope angle $\alpha \left( x \right)=-\mbox{arctan}\,\,\left[ {{\phi }'(x)} 
\right]$ should monotonically decrease from some initial value $\alpha _0 $ 
to the final smaller one $\alpha _F $, so that the region$\,\alpha _F \le 
\alpha \le \alpha _0 $ is under consideration . This idea has been suggested 
by numerical calculations for the descent time along parabolas with 
different signs of curvature \cite{Kurilin2023}. This statement is also confirmed by the 
cycloid equation, which is the curve of the fastest descent when $\mu =0$. 
The parametric equations of the cycloid 
$$
\left\{ {\begin{array}{l}
 x(\tau )=C_0 \left( {\tau -\sin \tau } \right) \\
 y(\tau )=H-C_0 \left( {1-\cos \tau } \right) \\
 \end{array}} \right.,
$$
can be rewritten in a similar form:
\begin{equation}
\label{CycloidEqn}
\left\{ {\begin{array}{l}
 x(\alpha )=C_0  \left( {\pi -2\alpha -\sin 2\alpha } \right) \\
 y(\alpha )=H-C_0  \left( {1+\cos 2\alpha } \right) \\
 \end{array}} \right.,
\end{equation}
which corresponds to the choice $\psi \left( \alpha \right)\!=\!-\frac{\displaystyle 1}{\displaystyle 2}
\cos^2\alpha ,\,\, \alpha _0 =\pi / 2,\,\, C_0 =C / 8 \,$ 
in formulas (\ref{XY-alpha}). In this 
case, the ``time'' parameter $\tau $ in the cycloid is related to the slope 
of the cycloid graph $\alpha \left( x \right)$ by simple relationship:
\begin{equation}
\label{Angles16}
\alpha =\frac{\pi -\tau }{2}.
\end{equation}
The limiting value of $\tau $-variable included in the cycloid equations 
(\ref{CycloidEqn}), (i.e. $0\le \tau \le \tau _0 )$ is associated with the final steepness 
of the cycloid $\alpha _F $ at the point $x_F =L$
$$
\alpha _F =\frac{\pi -\tau _0 }{2}.
$$
Exact values of parameters $C_0, \tau _0$ for the cycloid equations 
(\ref{CycloidEqn}) can be calculated from following system of equations which relates them 
to the Cartesian coordinates of the beginning and ending points $A(0;H)$ and 
$B(L;0)$:
\begin{equation}
\label{LH-tau}
\left\{ {\begin{array}{l}
 L=C_0  \left( {\tau _0 -\sin \tau _0 } \right) \\ 
 H=C_0  \left( {1-\cos \tau _0 } \right) \\ 
 \end{array}} \right., 
\end{equation}
The solution to this system which we name the boundary conditions can be 
obtained only numerically. For example, the choice of the initial data 
$H=5\,m,\,\,L=4\,m,$ corresponds to the following values in the SI 
system \cite{Kurilin2016}:
\begin{equation}
\label{Ctau0}
C_0 =3.40517104\,\mbox{\it m},\quad \tau _0 =2.0582244.
\end{equation}
When friction forces are taken into account the parameters $C, \alpha _F$ 
included in the brachistochrone equations (\ref{XY-alpha}) must also obey some boundary 
conditions similar to that of transcendental system (\ref{LH-tau}). Thus, we arrive to 
the following equations:
\begin{equation}
\label{Lpsi}
L=C \int\limits_{\alpha _0 }^{\alpha _F } {\psi (\beta )} \,d\beta .
\end{equation}
\begin{equation}
\label{Hpsi}
H=C \int\limits_{\alpha _0 }^{\alpha _F } {\psi (\beta )\, 
\mbox{tan}\beta } \,d\beta .
\end{equation}
Substituting parametric equations of the supposed brachistochrone (\ref{XY-alpha}) into 
(\ref{Wsol11}) -- (\ref{Time13}) we get the expression for the total descent time, written as an 
integral of some unknown function $\psi (\beta )$:
\begin{equation}
\label{Time-mu}
T(\mu )=\sqrt {\frac{C}{2g}} \int\limits_{\alpha _0 }^{\alpha _F } 
{\frac{e^{-\mu \beta }}{\cos \beta }\frac{\psi (\beta )}{\sqrt {\Phi (\beta 
)} }} \,d\beta .
\end{equation}
In this case, the ``acceleration history'' function (\ref{Kint12}) transforms with 
constant multiplier $\sqrt {2gC} $ to the dimensionless integral of the same 
function $\psi (\beta )$, which is also included in eq. (\ref{Time-mu}) through the 
following modification
\begin{equation}
\label{FunPhi}
\Phi \left( \alpha \right)=\int\limits_{\alpha _0 }^\alpha {\left( 
{\mbox{tan}\beta -\mu } \right)e^{-2\mu \beta }\, \psi (\beta )\,d\beta }.
\end{equation}
From a mathematical point of view, the problem of finding the fastest 
descent curve (brachistochrone with friction) has been reduced to finding 
the minimum value of the functional (\ref{Time-mu}) by choosing a ``suitable'' function 
$\psi (\beta )$ and the corresponding parameters $C, \alpha _F, \, \alpha _0 $ 
with restrictions (\ref{Lpsi}), (\ref{Hpsi}). Now we can talk more specifically about 
different approaches to solving the problem of brachistochrone with Coulomb 
friction. In the work \cite{Ashby} and subsequent publications, the problem of 
finding the extremum of functional (\ref{Time-mu}) was combined with boundary 
conditions (\ref{Lpsi})~-~(\ref{Hpsi}), which led to very complex equations. The authors 
were unable to solve these equations analytically and find adequate formulas 
for brachistochrone with friction. In this article, we propose first to find 
the shape of the extremal curves for functional (\ref{Time-mu}), and then, by choosing 
the optimal parameters $C, \alpha _F,  \alpha _0$ ensure that boundary 
conditions (\ref{Lpsi})~-~(\ref{Hpsi}) are met. This idea was suggested by analytical and 
numerical calculations of the sliding time along a cycloid in the presence 
of dry friction force \cite{Kurilin-Arxiv}. Note that at zero friction $\mu =0$, if the 
cycloid equations (\ref{CycloidEqn}) are substituted into (\ref{Time-mu}), (\ref{FunPhi}) then the integrand in 
formula (\ref{Time-mu}) becomes a constant and does not depend on the inclination angle 
of the trajectory $\beta $ at a given point. Thus, the time of descent is 
determined only by the parameters $C,\, \alpha _0,\, \alpha _F$:
\begin{equation}
\label{Time0}
T(0)=\sqrt {\frac{C}{2g}} \left( {\alpha _0 -\alpha _F } \right)=\tau 
_0 \sqrt {\frac{C_0 }{g}}. 
\end{equation}
Let us use this condition as a criterion for finding the minimum value of 
the functional (\ref{Time-mu}), because if the integrand is a constant then its 
variation will be equal to zero as must be for the extremum value. We 
suppose that the shape of the desired brachistochrone is determined by the 
condition:
\begin{equation}
\label{Ansatz}
\frac{e^{-\mu \beta }}{\cos \beta }\frac{\psi (\beta )}{\sqrt {\Phi (\beta 
)} }\equiv -1.
\end{equation}
The choice of the constant number "-1" in the right side of formula (\ref{Ansatz}) is 
quite arbitrary and can always be changed by redefining the scale factor 
$C$. Then the time of descent along the optimal curve will depend only on 
the scale parameter $C(\mu )$ and the inclination angles of the curve at the 
initial and final points of the trajectory $\alpha _0 (\mu )$, $\alpha _F 
(\mu )$.
\begin{equation}
\label{TimeMu2}
T(\mu )=\sqrt {\frac{C}{2g}} \, \int\limits_{\alpha _0 }^{\alpha _F } 
{\frac{e^{-\mu \beta }}{\cos \beta }\frac{\psi (\beta )}{\sqrt {\Phi (\beta 
)} }} \,d\beta =\sqrt {\frac{C(\mu )}{2g}} \, \left[ {\alpha _0 (\mu 
)-\alpha _F (\mu )} \right]
\end{equation}
Relationship (\ref{Ansatz}) is equivalent to the following integral equation:
\begin{equation}
\label{EqnPhi2}
\Phi \left( \alpha \right)=\int\limits_{\alpha _0 }^\alpha {\left( 
{\mbox{tan}\beta -\mu } \right)e^{-2\mu \beta }\, \psi (\beta )\,d\beta } 
=\frac{e^{-2\mu \alpha }}{\cos ^2\alpha }\psi ^2\left( \alpha \right).
\end{equation}
We differentiate both sides of equation (\ref{EqnPhi2}) with respect to the variable 
$\alpha$ and get:
\begin{equation}
\label{EqnPsi}
{\psi }'\left( \alpha \right)+\left( {\mbox{tan}\alpha -\mu } \right) 
\psi \left( \alpha \right)=\frac{1}{2}\cos ^2\alpha \left( 
{\mbox{tan}\alpha -\mu } \right).
\end{equation}
Solving this first-order linear differential equation with the initial 
condition $\psi \left( {\alpha _0} \right)=0$, (which corresponds to the 
zero initial speed $\upsilon (0)=0)$, we obtain the formula:
\begin{equation}
\label{psiA}
\psi \left( \alpha \right)=\frac{\cos \alpha }{2\left( {1+\mu ^2} 
\right)}\left[ {e^{\mu (\alpha -\alpha _0 )}\left( {(1-\mu ^2)\cos \alpha _0 
+2\mu \sin \alpha _0 } \right)-(1-\mu ^2)\cos \alpha -2\mu \sin \alpha } 
\right]
\end{equation}
Note that the choice $\alpha _0 =\frac{\displaystyle\pi }{\displaystyle 2}$  
which was made in papers \cite{Ashby,Sumbatov} because 
$\psi \left( {\frac{\displaystyle\pi }{\displaystyle 2}} \right)=0$ is not 
justified and erroneous from our point of view. The initial inclination 
angle of the trajectory $\alpha _0 $ is also determined by the friction 
coefficient $\mu $ and will be calculated below. Substituting (\ref{psiA}) into 
formulas (\ref{XY-alpha}) we obtain a parametric form of the desired brachistochrone 
with Coulomb friction forces:
\begin{equation}
\label{Bra-mu}
x\left( \alpha \right)=C\cdot FX\left( {\alpha ,\alpha _0 } \right),
\quad
y\left( \alpha \right)=H-C\cdot FY\left( {\alpha ,\alpha _0 } \right),
\end{equation}
where
\begin{eqnarray}
\label{FX}
 FX\left( {\alpha ,\alpha _0 } \right)=\frac{e^{\mu (\alpha -\alpha _0
)}}{2\left( {1+\mu ^2} \right)^2}\left[ {(1-\mu ^2)\cos \alpha _0 +2\mu \sin
\alpha _0 } \right]\left( {\sin \alpha +\mu \cos \alpha } \right)+ \nonumber \\
 +\frac{\mu \cdot \cos \left( {2\alpha } \right)}{4\left( {1+\mu ^2}
\right)}-\frac{(1-\mu ^2)}{8\left( {1+\mu ^2} \right)}\left( {2\alpha
-2\alpha _0 +\sin \left( {2\alpha } \right)} \right)-\frac{\mu (3-\mu
^2)}{4\left( {1+\mu ^2} \right)^2}-\frac{1}{8}\sin \left( {2\alpha _0 }
\right) ,
\end{eqnarray}

\begin{eqnarray}
\label{FY}
 FY\left( {\alpha ,\alpha _0 } \right)=\frac{e^{\mu (\alpha -\alpha _0
)}}{2\left( {1+\mu ^2} \right)^2}\left[ {(1-\mu ^2)\cos \alpha _0 +2\mu \sin
\alpha _0 } \right]\left( {\mu \sin \alpha -\cos \alpha } \right)- \nonumber \\
 -\frac{\mu }{4\left( {1+\mu ^2} \right)}\left( {2\alpha -2\alpha _0 -\sin
\left( {2\alpha } \right)} \right)+\frac{(1-\mu ^2)}{8\left( {1+\mu ^2}
\right)}\cos \left( {2\alpha } \right)+\frac{(1-3\mu ^2)}{4\left( {1+\mu ^2}
\right)^2}+\frac{1}{8}\cos \left( {2\alpha _0 } \right) .
 \end{eqnarray}
Since the formulas (\ref{FX}), (\ref{FY}) turned out to be quite cumbersome, let us 
rewrite them in trigonometric form, using the critical friction angle 
$\gamma =\mbox{arctan}\left( \mu \right)$. As known, downward sliding 
without an initial speed is possible only if the inclination angle of the 
trajectory is greater than it ( $\alpha _0 >\gamma )$.
\begin{equation}
\label{Mu-gamma}
\sin \gamma =\frac{\mu }{\sqrt {1+\mu ^2} }
\quad
\cos \gamma =\frac{1}{\sqrt {1+\mu ^2} }\,
\end{equation}
In this form of notation, equations (\ref{psiA}) -- (\ref{FY}) take the form:
\begin{equation}
\label{psi2}
\psi \left( \alpha \right)=\frac{1}{2}\cos \alpha \left[ {e^{\mu (\alpha
-\alpha _0 )}\cos \left( {\alpha _0 -2\gamma } \right)-\cos \left( {\alpha
-2\gamma } \right)} \right],
\end{equation}

\begin{eqnarray}
\label{FX2}
 FX\left( {\alpha ,\alpha _0 } \right)=\frac{1}{2}e^{\mu (\alpha -\alpha _0
)}\cos \gamma \cdot \cos \left( {\alpha _0 -2\gamma } \right)\cdot \sin
\left( {\alpha +\gamma } \right)-\frac{1}{4}\sin \alpha \cdot \cos \left(
{\alpha -2\gamma } \right)- \nonumber \\
 -\frac{1}{8}\sin (2\alpha _0 )-\frac{1}{8}\sin (4\gamma )+\frac{1}{4}\left(
{\alpha _0 -\alpha } \right)\cdot \cos \left( {2\gamma } \right),
\end{eqnarray}

\begin{eqnarray}
\label{FY2}
 FY\left( {\alpha ,\alpha _0 } \right)=-\frac{1}{2}e^{\mu (\alpha -\alpha _0
)}\cos \gamma \cdot \cos \left( {\alpha _0 -2\gamma } \right)\cdot \cos
\left( {\alpha +\gamma } \right)+\frac{1}{4}\cos \alpha \cdot \cos \left(
{\alpha -2\gamma } \right)+ \nonumber \\
 +\frac{1}{8}\cos (2\alpha _0 )+\frac{1}{8}\cos (4\gamma )+\frac{1}{4}\left(
{\alpha _0 -\alpha } \right)\cdot \sin \left( {2\gamma } \right).
\end{eqnarray}
Now we proceed to finding other free parameters $C$, $\alpha _0 $, $\alpha 
_F $ of the brachistochrone minimizing the value of the functional (\ref{Time-mu}) with 
boundary conditions (\ref{Lpsi}), (\ref{Hpsi}). Once the shape of the optimal curve is found 
the problem reduces to calculate the extremum value of the function (\ref{TimeMu2}):
\begin{equation}
\label{Time3mu}
T(C,\alpha _0 ,\alpha _F ,\mu )=\sqrt {\frac{C(\mu )}{2g}} \cdot \left[ 
{\alpha _0 (\mu )-\alpha _F (\mu )} \right]
\end{equation}
with boundary conditions that follow from (\ref{Lpsi}), (\ref{Hpsi}):
\begin{equation}
\label{HLeqn}
L=C\cdot FX\left( {\alpha { }_F,\alpha _0 } \right),
\quad
H=C\cdot FY\left( {\alpha _F ,\alpha _0 } \right).
\end{equation}
Let us use the standard Lagrange method with multipliers 
$\lambda _1,\,\lambda _2$ and consider the auxiliary Lagrange function combining 
formula (\ref{Time3mu}) and restrictions (\ref{HLeqn}):
\begin{equation}
\label{Lagrange38}
L(C,\alpha _0 ,\alpha _F )=\sqrt {\frac{C}{2g}} \cdot \left( {\alpha _0 
-\alpha _F } \right)+\lambda _1 \left[ {C\cdot FX\left( {\alpha { }_F,\alpha 
_0 } \right)-L} \right]+\,\lambda _2 \left[ {C\cdot FY\left( {\alpha _F 
,\alpha _0 } \right)-H} \right]
\end{equation}
Partial derivative of the Lagrange function (\ref{Lagrange38}) with respect to $C$ 
variable gives the first equation for the extremum point:
\begin{equation}
\label{Lambda39}
\lambda _1 \cdot FX\left( {\alpha { }_F,\alpha _0 } \right)+\,\lambda _2 
\cdot FY\left( {\alpha { }_F,\alpha _0 } \right)=-\,\,\frac{\left( {\alpha 
_0 -\alpha { }_F} \right)}{2\sqrt {2gC} }.
\end{equation}
Other partial derivatives with angles $\alpha _0 $, $\alpha _F $ produce a 
system of linear equations:
\begin{equation}
\label{sys22}
\left\{ {\begin{array}{l}
a_{11}\cdot\lambda_1 +\, a_{12}\cdot\lambda_2  =-\,\frac{\displaystyle 1}{\displaystyle \sqrt{2gC}} \\
a_{21}\cdot\lambda_1 +\, a_{22}\cdot\lambda_2  =\,\,\,\frac{\displaystyle 1}{\displaystyle \sqrt{2gC}}, \\
\end{array}}
\right. 
\end{equation}
where the elements of the system matrix have the form:
\begin{equation}
\label{sysElements41}
\begin{array}{l}
 a_{11} =\frac{\displaystyle \partial FX\left( {\alpha { }_F,\alpha _0 } \right)}
 {\displaystyle \partial \alpha _0 };\qquad 
a_{12} =\frac{\displaystyle\partial FY\left( {\alpha { }_F,\alpha_0 } \right)}
{\displaystyle\partial \alpha _0 };\\
\\
 a_{21} =\frac{\displaystyle \partial FX\left( {\alpha { }_F,\alpha _0 } \right)}
{\displaystyle\partial \alpha _F };\qquad 
 a_{22} =\frac{\displaystyle\partial FY\left( {\alpha { }_F,\alpha_0 } \right)}
{\displaystyle\partial \alpha _F }. \\
\end{array}
\end{equation}
Solving the system (\ref{sys22}) we find the Lagrange multipliers 
$\lambda _1,\,\lambda _2 $:
\begin{equation}
\label{Lambda42}
\begin{array}{l}
\lambda _1 =\frac{\displaystyle1}{\displaystyle \Delta \sqrt {2gC} }
\left( {\displaystyle\frac{\partial FY\left({\alpha { }_F,\alpha _0 } \right)}
{\displaystyle\partial \alpha _0 }+\frac{\displaystyle \partial FY\left( {\alpha { }_F,\alpha _0 } \right)}
{\displaystyle\partial \alpha _F }} \right),\\
\\
\lambda _2 =\frac{\displaystyle 1}{\displaystyle\Delta \sqrt {2gC} }
\left( {\frac{\displaystyle\partial FX\left( {\alpha { }_F,\alpha _0  } \right)}
{\displaystyle \partial \alpha _0 }+\frac{\displaystyle\partial FX\left( {\alpha { }_F,\alpha _0 } \right)}
{\displaystyle\partial \alpha _F }} \right) ,
\end{array}
\end{equation}
where $\Delta$ is the main determinant of the system:
\begin{equation}
\label{DetSys43}
\Delta =\left| {{\begin{array}{*{20}c}
 {a_{11} } \hfill & {a_{12} } \hfill \\
 {a_{21} } \hfill & {a_{22} } \hfill \\
\end{array} }} \right| 
=\frac{\partial FX\left( {\alpha { }_F,\alpha_0 }
\right)}{\partial \alpha _0 } \cdot \frac{\partial FY\left( {\alpha { }_F,\alpha _0
} \right)}{\partial \alpha _F }-\frac{\partial FX\left( {\alpha { }_F,\alpha
_0 } \right)}{\partial \alpha _F }\cdot \frac{\partial FY\left( {\alpha {
}_F,\alpha _0 } \right)}{\partial \alpha _0 }.
\end{equation}
Substituting the found solutions (\ref{Lambda42}) into the first extremum equation (\ref{Lambda39}) 
we find the sought interrelation between the initial $\alpha _0$ and final 
$\alpha _F$ slope of the brachistochrone graph with friction coefficient 
$\mu =\tan \left( \gamma \right)$.
\begin{equation}
\label{GenEqn44}
\left( {\frac{\partial FY}{\partial \alpha _0 }+\frac{\partial FY}{\partial
\alpha _F }} \right)FX\left( {\alpha { }_F,\alpha _0 } \right)-\left(
{\frac{\partial FX}{\partial \alpha _0 }+\frac{\partial FX}{\partial \alpha
_F }} \right)FY\left( {\alpha { }_F,\alpha _0 } \right)=\frac{\Delta
}{2}\left( {\alpha _0 -\alpha { }_F} \right).
\end{equation}
The determinant of the system of linear equations (\ref{sys22}) after substitution 
formulas (\ref{FX2}), (\ref{FY2}) into equation (\ref{DetSys43}) takes the form
\begin{eqnarray}
\label{DetMod45}
\Delta =\frac{1}{4}\sin\left( {\gamma -\alpha _0}\right) \nonumber\times \\
\times \left[ {e^{\mu
(\alpha _F -\alpha _0 )}\cos \gamma -\cos \left( {\alpha _0 -\alpha _F
+\gamma } \right)} \right] \left[ {e^{\mu (\alpha _F -\alpha _0 )}\cos \left(
{\alpha _0 -2\gamma } \right)-\cos \left( {\alpha_F -2\gamma }\right)}
\right]\!.
\end{eqnarray}
The required relationship between $\alpha _0 ,\alpha _F $ and $\mu $ which 
we obtain from eq. (\ref{GenEqn44}) with the aid of (\ref{FX2}), (\ref{FY2}), (\ref{DetMod45}) after simple but 
rather cumbersome transformations can be written as follows
\begin{eqnarray}
\label{eqnAlpha}
 \frac{1}{4}e^{2\mu (\alpha _F -\alpha _0 )}\cos^2\gamma \cdot \cos^2( \alpha _0 -2\gamma)-  \nonumber\\
-\frac{1}{4} e^{\mu (\alpha _F -\alpha _0 )}\cos\gamma 
 \left[ {\cos (\alpha _0 -\alpha _F +\gamma )+\frac{1}{2}\cos ( \alpha _0 +\alpha _F -5\gamma ) 
 +\frac{1}{2}\cos ( \alpha _0 -\alpha _F ) \cdot \cos ( 2\alpha _0 -3\gamma ) }\right] +  \nonumber\\
 +\frac{1}{8}\cos (\alpha _0 -\alpha _F +\gamma )
 \left[ {\cos (\alpha _0 -\alpha _F +\gamma )+\cos \gamma \cdot 
 \cos (\alpha _0 +\alpha _F -4\gamma )} \right]= \\
 =-\frac{1}{8}\left( {\alpha _0 -\alpha _F } \right)e^{2\mu (\alpha _F
-\alpha _0 )}\cos \gamma \cdot \cos \left( {\alpha _0 -2\gamma }
\right)\cdot \sin \left( {\alpha _0 -\gamma } \right)+ \nonumber\\
 +\frac{1}{16}\left( {\alpha _0 -\alpha _F } \right)e^{\mu (\alpha _F
-\alpha _0 )}\left[ {\sin \left( {\alpha _0 -\alpha _F +2\gamma }
\right)+\sin \left( {\alpha _0 -\alpha _F } \right)\cdot \cos \left(
{2\alpha _0 -2\gamma } \right)} \right]+ \nonumber\\
 +\frac{1}{8}\left( {\alpha _0 -\alpha _F } \right)\cos \left( {\alpha _0
-\alpha _F +\gamma } \right)\sin \left( {\alpha _F -2\gamma } \right)\cos
\left( {\alpha _0 -\gamma } \right) \nonumber
\end{eqnarray}
In spite of quite complicated form of this equation, it can be easily 
analyzed by computer methods. For numerical calculations, the ``Mathcad'' 
program was used, the worksheet code in which can be viewed in the Internet 
via a link to the author's website \cite{Kurilin2015}. It is also noteworthy that at zero 
friction $\mu =0$ the equation (\ref{eqnAlpha}) takes the simplest form:
\begin{equation}
\label{eqnmu0}
\cos \alpha _0 \left( {\cos \alpha _0 -\cos \alpha _F } \right)\sin \left( 
{\alpha _0 -\alpha _F } \right)\left[ {\tan \left( {\frac{\alpha _0 -\alpha 
_F }{2}} \right)-\frac{\alpha _0 -\alpha _F }{2}} \right]=0,
\end{equation}
from which it follows that $\alpha _0 =\frac{\displaystyle \pi }{\displaystyle 2}$, as is true for the 
cycloid (\ref{CycloidEqn}). However, when friction forces are taken into account, the 
initial and final inclination angles of the brachistochrone graph $\alpha _0 
$,$\alpha _F$ are related to the sliding friction coefficient $\mu $ and 
the parameters of the problem $H,L$ by a combined system consisting of 
equation (\ref{eqnAlpha}) and the formula resulting from relations (\ref{HLeqn}):
\begin{equation}
\label{eqnAlpha2}
\frac{FY\left( {\alpha _F ,\alpha _0 } \right)}{FX\left( {\alpha _F ,\alpha 
_0 } \right)}=\frac{H}{L}.
\end{equation}

\section{NUMERICAL ANALYSIS}

\begin{figure}[htbp]
\centerline{\includegraphics[width=145mm]{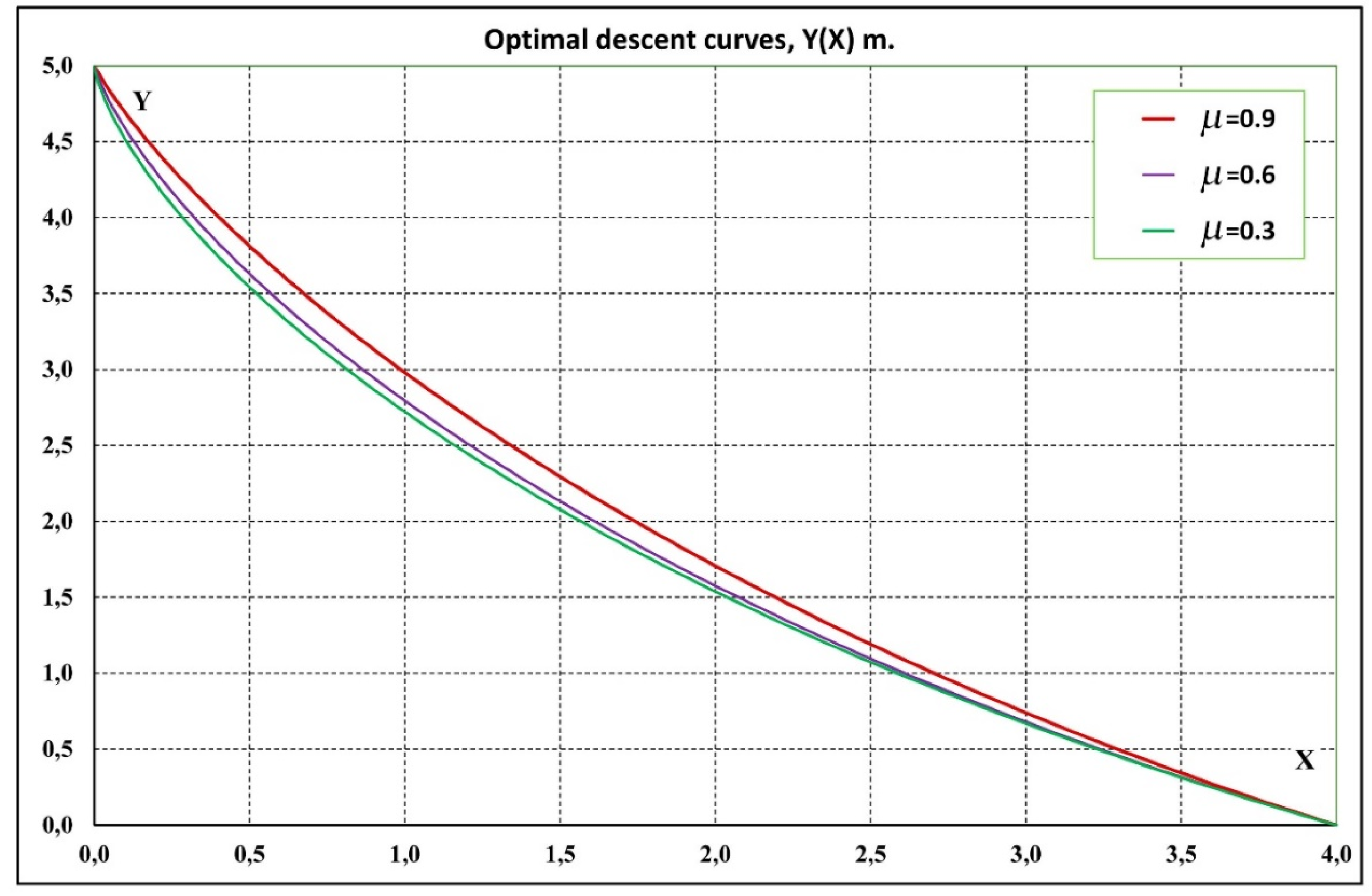}}
\caption[]{Shape of brachistochrones for different values of the 
friction coefficient $\mu $.}
\label{Curves-3}
\end{figure}

\begin{figure}[htbp]
\centerline{\includegraphics[width=140mm]{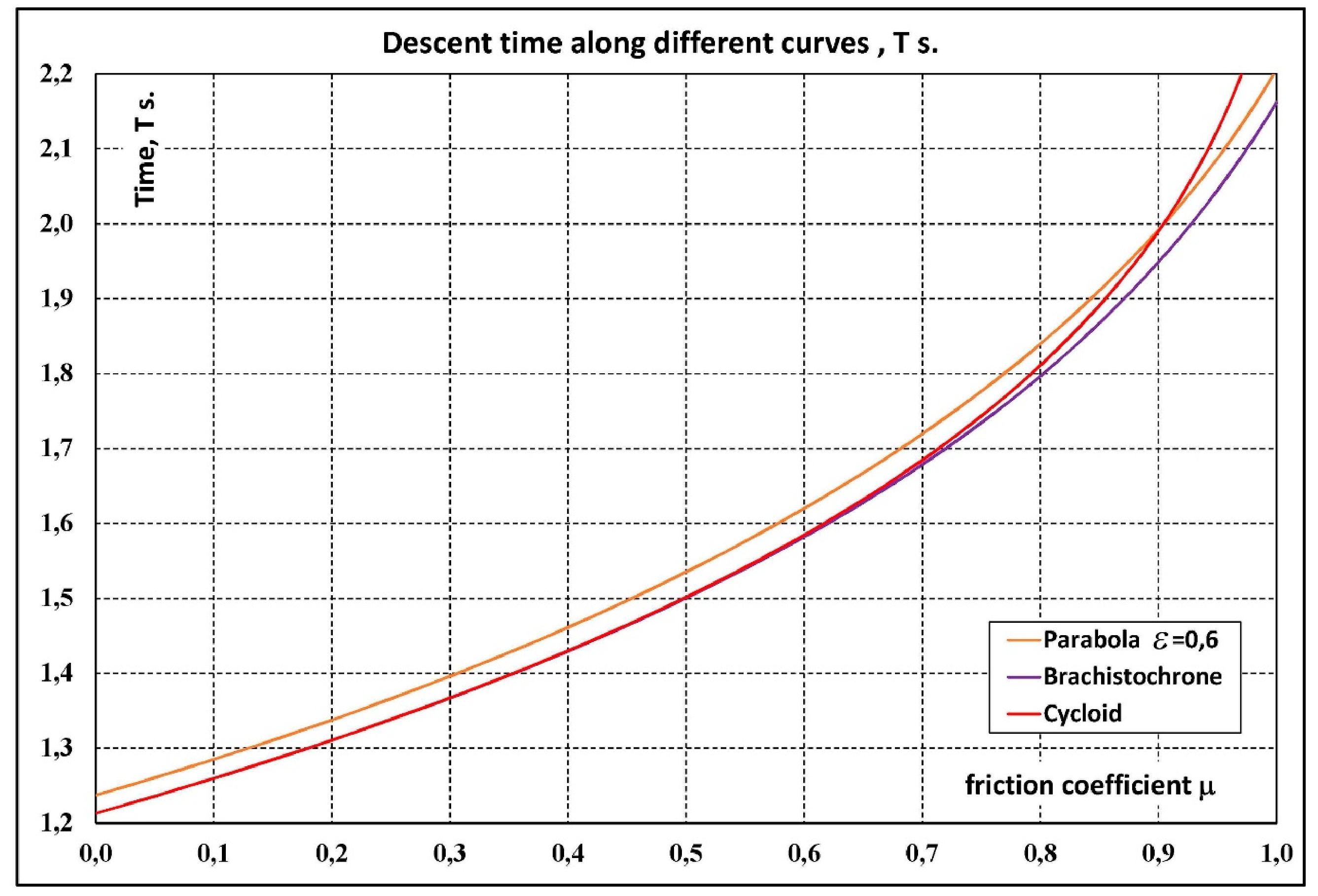}}
\caption[]{The descent time along different curves as functions 
of the sliding friction coefficient $\mu $.}
\label{Times-mu}
\end{figure}

Let us now carry out a numerical analysis of the obtained expressions, 
comparing the time of movement along the found brachistochrone and the 
cycloid at different values of the friction coefficient $0\le \mu \le 1$. We 
choose the parameters of the problem in accordance with the previously 
considered slide profile (\ref{Ctau0}):
\begin{equation}
\label{HillData}
H=5\,m,\quad L=4\,m,\quad g=9,81\,ms^{-2}.
\end{equation}
All reasoning carried out above shows that there is no universal curve that 
provides the minimum descent time for an arbitrary value of the friction 
coefficient $\mu$. We can only talk about a family of curves, the shape of 
which also depends on the friction coefficient. Figure \ref{Curves-3}  shows three such 
curves given by formulas (\ref{Bra-mu}), (\ref{FX2}), (\ref{FY2}) for some different values of the 
friction coefficient $\, \mu =0.3;\, \mu =0.6;\, \mu =0.9.\,$
For the given parameters of the problem (\ref{HillData}), these curves do not differ 
very much from each other and are externally similar to the cycloid \cite{Kurilin2016}, 
into which they degenerate at $\mu \to 0$. The descent time along the family 
of these curves (proposed brachistochrones) (\ref{FX2}), (\ref{FY2}) for different values 
of the friction coefficient is shown in Fig. \ref{Times-mu}. Here, for comparison, we 
also depict the time of descent along the cycloid (\ref{CycloidEqn}) and along the 
parabola (\ref{ParabolaEQ}) with the curvature parameter $\varepsilon =0.6$, which was 
also considered in our previous work \cite{Kurilin-Arxiv}.
\begin{equation}
\label{ParabolaEQ}
\phi (x)=\frac{\kappa \varepsilon }{L}x^2-\kappa \left( {1+\varepsilon } 
\right)x+H,
\quad
\kappa =\frac{H}{L}.
\end{equation}

\begin{figure}[htbp]
\centerline{\includegraphics[width=140mm]{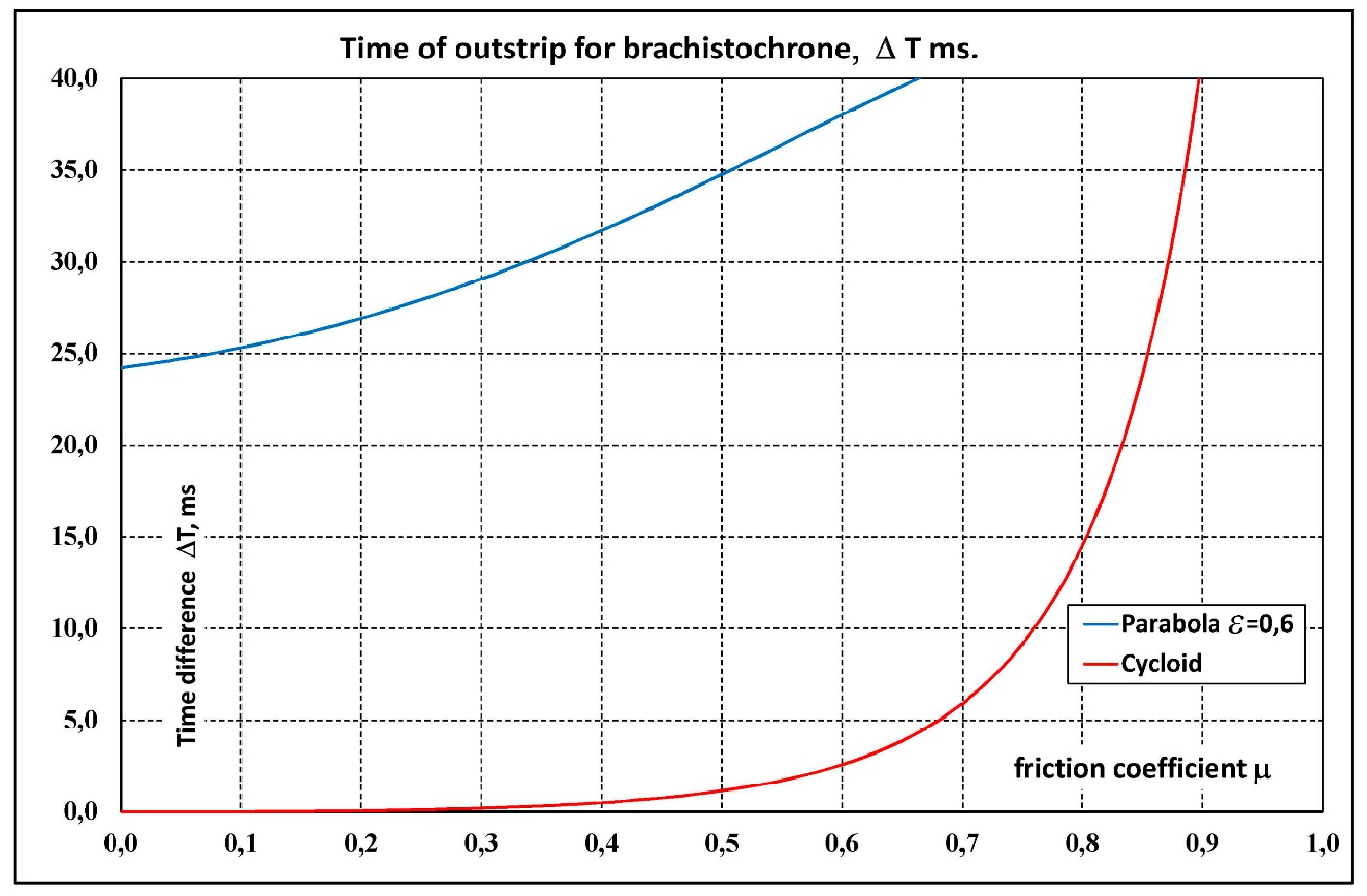}}
\caption{The difference in the descent time along the cycloid (\ref{CycloidEqn}) 
and along the parabola (\ref{ParabolaEQ}) compared with the time of movement 
on the brachistochrones (\ref{Bra-mu}).}
\label{Delta-times}
\end{figure}

\begin{figure}[htbp]
\centerline{\includegraphics[width=150 mm]{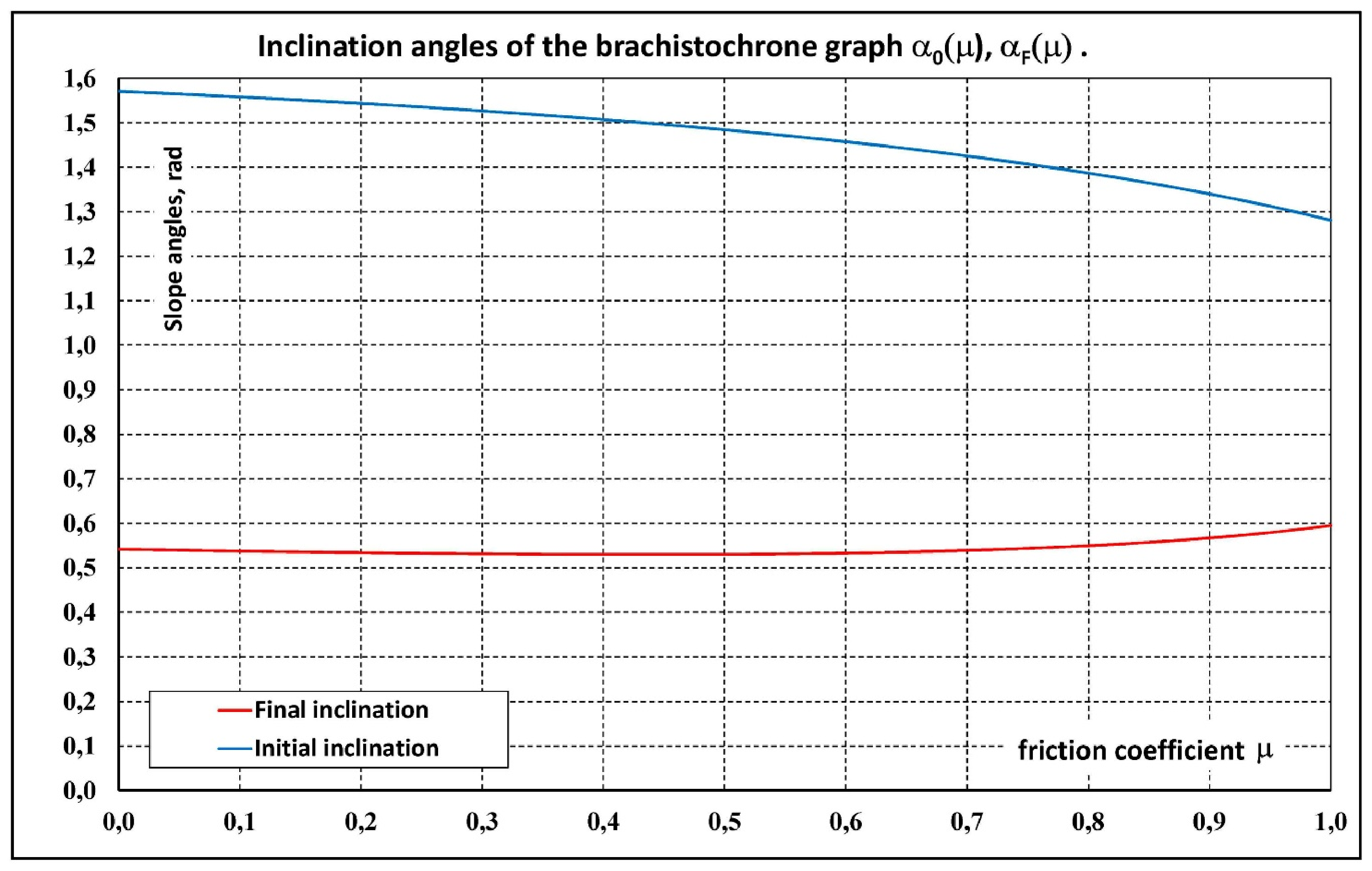}}
\caption{The dependence of the initial and final inclination angles 
of the brachistochrone curve family (\ref{Bra-mu}) on the friction coefficient.}
\label{Alphas}
\end{figure}

These graphs show that up to the values, $\mu \approx 0.7$ the differences 
between the family of the curves (\ref{Bra-mu}), (\ref{FX2}), (\ref{FY2}) and the cycloid (\ref{CycloidEqn}) are 
practically unnoticeable. The particle on the proposed brachistochrone 
outstrips the one on the cycloid in descent time about some milliseconds, 
which is clearly illustrated in Fig. \ref{Delta-times}. So, numerical calculations show that 
the family of the above-mentioned curves (\ref{Bra-mu}) -- (\ref{FY}) can claim to be 
the brachistochrone for a given sliding friction coefficient. Numerical 
experiments with other initial data of the problem, different from (\ref{HillData}), 
show that this statement remains valid and the family of curves (\ref{FX2}), (\ref{FY2}) 
provides the minimum descent time for a given $\mu $ value. It is curious to 
state that in the case $H=1\,m,\,\,L=6\,m,$ which was analyzed in \cite{Ashby,Sumbatov} a 
particle on the cycloid is not able to finish and stops somewhere halfway if 
the friction coefficient is greater than the value $\mu \ge 0.10$. This 
occurs because of the condition $\alpha _0 =\frac{\displaystyle \pi }{\displaystyle 2}$ 
implying fast acceleration and great increase of the friction forces on the 
rather curved trajectory. At the same time the particle on the 
brachistochrone (\ref{Bra-mu})-(\ref{FY}) with $\alpha _0 <\frac{\displaystyle \pi }{\displaystyle 2}$ 
is able to slide 
to the finish line $x_F =L$ up to the values $\mu \le 0.16$.
Numerical analysis of formulas (\ref{eqnAlpha}), (\ref{eqnAlpha2}) allows us to trace the dependence 
of the initial $\alpha _0$ and final $\alpha _F $ inclination angles of the 
brachistochrone graph (\ref{Bra-mu}) on the friction coefficient $\mu$. The 
calculation results are shown in Fig.\ref{Alphas}.
As can be seen from the graphs in Fig.\ref{Alphas}, the initial slope angle $\alpha _0 
(\mu )$ monotonously decreases from value $\alpha _0 (0)=1.5708$ to value 
$\alpha _0 (1.0)=1.2811$ with the increase of sliding friction coefficient 
$\mu $, i.e. changes by approximately 18{\%}. As for the final slope of the 
brachistochrone graph $\alpha _F (\mu )$, it increases by approximately 
9{\%} from the value $\alpha _F (0)=0.5417$ to $\alpha _F (1.0)=0.5947$. 
Thus, with an increase in the sliding friction coefficient, the 
brachistochrone graph straightens more and more and the radius of the 
average curvature of the trajectory becomes larger. This is clearly seen 
from Fig.\ref{Curves-3}  and has a simple physical explanation. The sliding friction 
forces decrease at high speed on a less concave trajectory, so the steep 
initial section of brachistochrone with fast acceleration smoothly 
transitions into a less concave section of almost straight finish.

\begin{figure}[htbp]
\centerline{\includegraphics[width=150 mm]{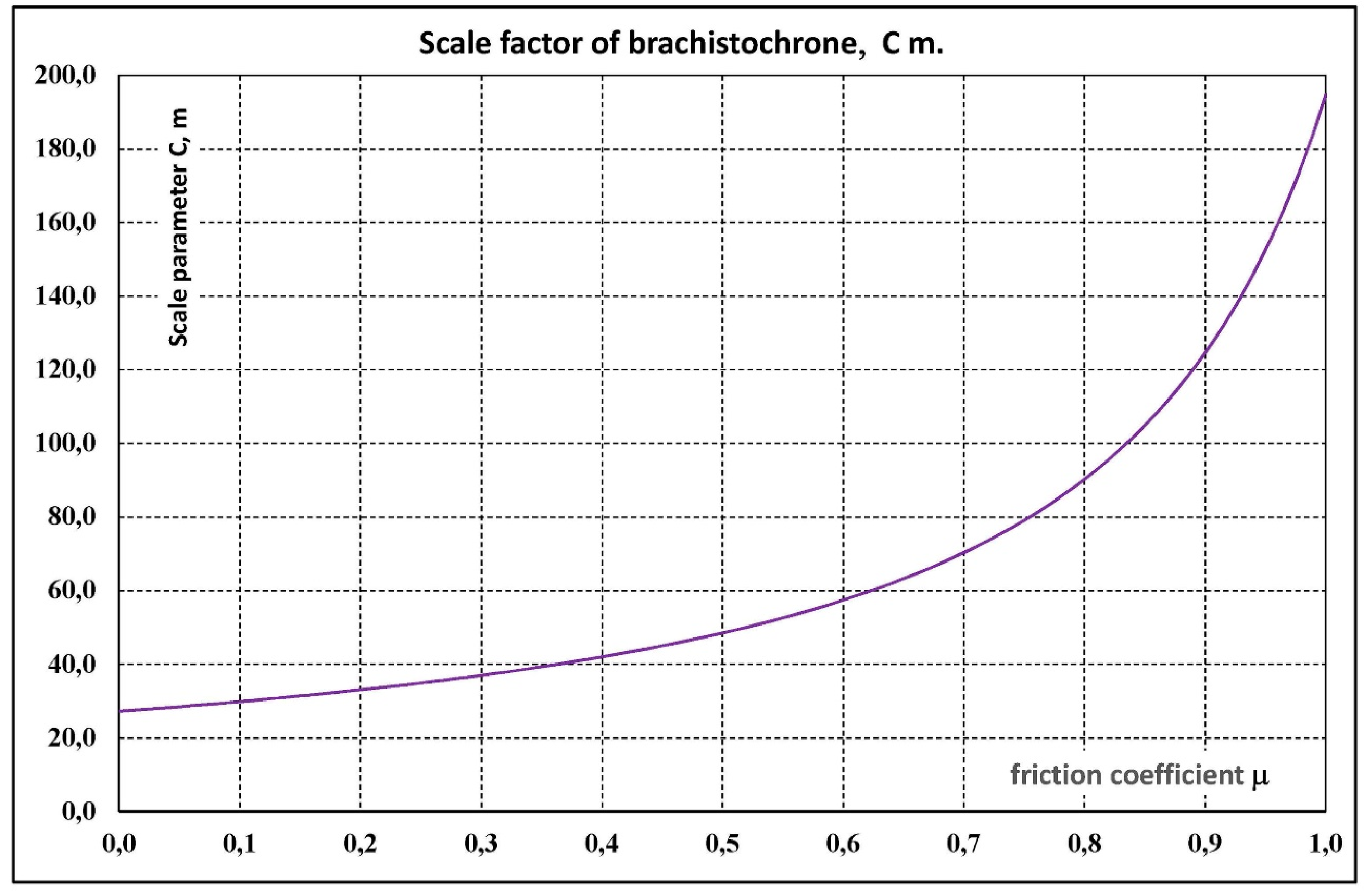}}
\caption{The dependence of the scale parameter $C(\mu )$ of the 
optimal curve trajectory (\ref{Bra-mu}), on the friction coefficient.}
\label{Cfactor}
\end{figure}

The scale parameter $C(\mu )$ of the fastest descent trajectory (\ref{Bra-mu}) can be 
calculated from formulas (\ref{HLeqn}) through the initial data of the problem (\ref{HillData}) 
and the found angles of initial $\alpha _0$ and final $\alpha _F $ 
steepness. This value turns out to be strongly dependent on the friction 
coefficient. Figure \ref{Cfactor} shows how the scale factor changes $C(\mu )$ with the 
increase of $\mu$. The initial value $C(0)=27.2414$ grows up more than 
seven times and reaches the magnitude $C(1.0)=194.6199$ at $\mu =1.0$.

Let us now discuss the geometric and physical meanings of function $\psi(\alpha )$ 
(\ref{psiA}), (\ref{psi2}) that defines the brachistochrone equation in 
parametric form (\ref{XY-alpha}). Using formula (\ref{rad4}), we can find the interrelation 
between this function and the curvature radius of the trajectory at a given 
location:
\begin{equation}
\label{RadPsi}
R(\alpha )=-\frac{C}{\cos \alpha }\psi (\alpha )
\end{equation}
For a cycloid, $\psi \left( \alpha \right)=-\frac{\displaystyle 1}{\displaystyle 2}\cos ^2\alpha $ and 
this formula takes the simplest form:
\begin{equation}
\label{RadCycloid}
R(\alpha )=\frac{C}{2}\cos \alpha 
\end{equation}
On the other hand, the function $\psi (\alpha )$ can be associated with the 
horizontal projection of the body velocity on curve (\ref{XY-alpha}), using formulas 
(\ref{Wfun}), (\ref{Wsol11}), (\ref{FunPhi})
\begin{equation}
\label{PsiVX}
\upsilon _x =\upsilon (\alpha )\cos \alpha =\sqrt {2gC} \cdot \cos \alpha 
\cdot e^{\mu \alpha }\sqrt {\Phi (\alpha )} .
\end{equation}
Taking into account the condition (\ref{EqnPhi2}), the last formula can be written in 
terms of the magnitude of the velocity vector in the form:
\begin{equation}
\label{PsiVelocity}
\upsilon (\alpha )=-\frac{\sqrt {2gC} }{\cos \alpha }\psi (\alpha ).
\end{equation}
On the cycloid (\ref{CycloidEqn}), the relation (\ref{PsiVelocity}) has the form of the well-known 
optical-geometric formula for the change in the speed of light when passing 
through a medium with a variable refractive index:
\begin{equation}
\label{BernoulliEq}
\upsilon (\alpha )=\sqrt {\frac{gC}{2}} \cos \alpha .
\end{equation}
As is known, it was this optical-mechanical analogy that led Bernoulli to 
his solution to the problem of the frictionless brachistochrone \cite{Tikhomirov}. An 
interesting relation follows from formulas (\ref{RadCycloid}) and (\ref{BernoulliEq})
\begin{equation}
\label{GladkovAssume}
\upsilon ^2(\alpha )=gR(\alpha )\cos \alpha .
\end{equation}
This formula was used in the papers \cite{Gladkov2020,Gladkov2023} and led the authors to the idea 
that even in the presence of the Coulomb friction forces, this equation 
connecting the speed of the particle with the curvature radius should be 
satisfied on the true brachistochrone. Note that our calculations indicate 
that relation (\ref{GladkovAssume}) is not satisfied on the brachistochrone (\ref{FX2}), (\ref{FY2}) and 
the following formula takes place instead:
\begin{equation}
\label{VaRadius}
\upsilon (\alpha )=\sqrt {\frac{2g}{C}} \cdot R(\alpha ).
\end{equation}
The physical meaning of this formulas (\ref{GladkovAssume}), (\ref{VaRadius}) is quite clear: the closer 
a particle on the brachistochrone moves to the finish line, the flatter its 
trajectory should be in order to reduce the value of the friction force 
associated with the curvature of the trajectory. For the cycloid, both 
formulas (\ref{GladkovAssume}), (\ref{VaRadius}) are satisfied, and only calculating the descent time 
makes it possible to find out which of the proposed curves can claim the 
title of brachistochrone with friction (see \cite{Gladkov2023}).

We use formulas (\ref{psi2}) and (\ref{PsiVelocity}) to calculate the velocity of a particle on the 
brachistochrone (\ref{Bra-mu}) - (\ref{FY}) and find:

\begin{equation}
\label{VelocityMu}
\upsilon (\alpha )=\sqrt {\frac{g C}{2}} \left[ {\cos \left( {\alpha
-2\gamma } \right)-e^{\mu (\alpha -\alpha _0 )}\cos \left( {\alpha _0
-2\gamma } \right)} \right].
\end{equation}

\begin{figure}[htbp]
\centerline{\includegraphics[width=150 mm]{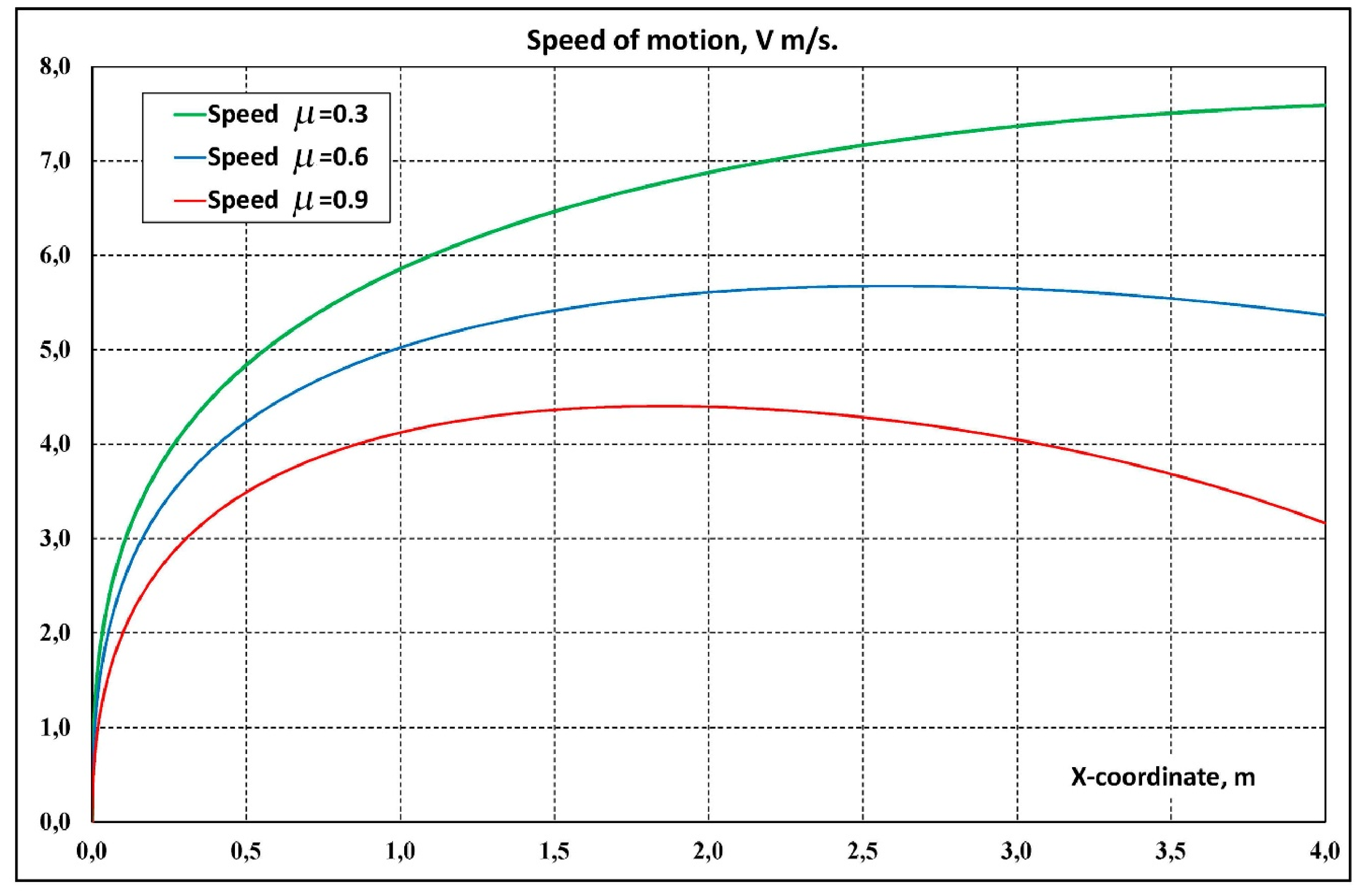}}
\caption{Change in the speed of a particle on the optimal curve (\ref{Bra-mu})
depending on the horizontal $x$-coordinate.}
\label{Speed-3mu}
\end{figure}

This relation generalizes Bernoulli's formula (\ref{BernoulliEq}) to the case of sliding 
with non-zero friction $\gamma =\mbox{arctan}(\mu )$ under the additional 
condition, that $\alpha _0 \left( \mu \right)\le \frac{\displaystyle \pi }{\displaystyle 2}$ 
and it is calculated separately. Figure \ref{Speed-3mu} illustrates how the speed on the generalized 
brachistochrone (\ref{Bra-mu}) -- (\ref{FY}) changes when the particle moves and horizontal 
$x$-coordinate increases from the start to final the value $x_F =L=4\,m$. 
Three curves with different friction coefficients as in Fig.~\ref{Curves-3} and with the 
same initial data (\ref{HillData}) were analyzed. As can be seen from the Fig.~\ref{Speed-3mu}, when 
$\mu \ge 0.5$ there appears sections with braking and velocity decrease. So 
the particle rapidly loses its kinetic energy and may even stop before 
reaching the finishing coordinate: $x_F =L$. The critical value of the 
friction coefficient at which the movement is still possible for this case 
is equal to $\mu _{\max} = 1.24$.

\section{CONCLUSION}

In this work, we studied motions of a small physical body (massive particle) 
along a family of some curves taking into account the Coulomb friction 
forces. It was found that these curves provide the minimum descent time of 
sliding with friction and can claim the title of brachistochrone 
generalizing the cycloid equation for this case. We have obtained new 
formulas for such optimal curves (\ref{Bra-mu}) - (\ref{FY}) which depend on the initial 
parameters of the problem and the friction coefficient $\mu $ in two ways 
explicitly and through three implicit monotonic functions $C(\mu ), \,
\alpha_F (\mu ), \, \alpha _0 (\mu )$, 
allowing only numerical analysis (\ref{HLeqn}), (\ref{eqnAlpha}), 
(\ref{eqnAlpha2}).  It is argued that the initial slope of the brachistochrone 
couldn't be equal to that one of the cycloid for all values of 
the friction coefficient $\mu $ and contrary to the widely used opinion  
the statement $\alpha _0 \left( \mu \right)\le \frac{\displaystyle \pi }{\displaystyle 2}$ is affirmed. 
The equations of the proposed brachistochrone with friction, make it 
possible to calculate the descent time at a given value of friction 
coefficient $\mu $ and compare it with the sliding time along other curves. 
Table \ref{tabTime} shows the results of such calculations for an isosceles profile 
$H=L$, where the descent time $T(\mu )$ (\ref{Time-mu}) is expressed in dimensionless 
relative units, which makes it possible to exclude the scale factor of such 
movement. For this, the following formula was used (see also \cite{Barsuk}):
\begin{equation}
\label{TimeTau}
\tau (\mu )=\frac{T(\mu )}{T_0 };\,
\quad
T_0 =\sqrt {\frac{L}{2g}} .
\end{equation}

\begin{table}[htbp]
\begin{center}
\begin{tabular}{|c|c|c|c|c|c|}
\hline
\textbf{$\mu $}&
\textbf{Circle}&
\textbf{Parabola}&
\textbf{Cycloid}&
\textbf{Curve}&
\textbf{Brachistochrone} \\
\textbf{}&
\textbf{}&
\textbf{ (\ref{ParabolaEQ}), $\varepsilon $=0,6}&
\textbf{(\ref{CycloidEqn})}&
\textbf{(17) from \cite{Gladkov2023}}&
\textbf{(\ref{Bra-mu})-(\ref{FY})} \\
\hline
0,0&
2,62206&
2,66174&
2,58190&
2,58190&
2,58190 \\
\hline
0,1&
2,75010&
2,78825&
2,70362&
2,70455&
2,70355 \\
\hline
0,2&
2,89716&
2,93102&
2,83995&
2,84519&
2,83961 \\
\hline
0,3&
3,07126&
3,09421&
2,99536&
3,01354&
2,99424 \\
\hline
0,4&
3,28824&
3,28381&
3,17686&
3,23534&
3,17372 \\
\hline
0,5&
3,58913&
3,50894&
3,39668&
3,83072&
3,38817 \\
\hline
0,6&
4,28398&
3,78472&
3,67962&
\textbf{-}&
3,65532 \\
\hline
0,7&
\textbf{-}&
4,13991&
4,09366&
\textbf{-}&
4,01019 \\
\hline
0,8&
\textbf{-}&
4,64714&
\textbf{-}&
\textbf{-}&
4,53711 \\
\hline
0,9&
\textbf{-}&
\textbf{-}&
\textbf{-}&
\textbf{-}&
5,53458 \\
\hline
\end{tabular}
\end{center}
\caption{Dimensionless time of descent $\tau (\mu )$ (\ref{TimeTau}) of a 
particle along different curves for different values of the friction 
coefficient $\mu $.}
\label{tabTime}
\end{table}

The dashes in this table indicate that the particle on this curve cannot 
reach the finish line and stops somewhere in the middle part of the 
trajectory, wasting the initial potential energy on the work against 
friction forces. The penultimate column of the table also shows calculations 
using formula (17) from the paper \cite{Gladkov2023}, in which the author used hypothesis 
(\ref{GladkovAssume}) for a brachistochrone with friction. Note that the formulas (19) from 
the same work, obtained using the law of energy conservation are incorrect 
in the presence of friction forces and therefore were not used in our 
calculations.

Comparing the data obtained, one can be convinced that curves (\ref{Bra-mu}) - (\ref{FY}) 
really provide the minimum descent time and the hypothesis (\ref{GladkovAssume}) is unfounded 
in the presence of friction forces. In addition, from formula (\ref{VaRadius}) it 
follows that
\begin{equation}
\label{AlphaTime60}
\frac{d\alpha (x(t))}{dt}=-\frac{\upsilon (\alpha )}{R(\alpha )}=-\sqrt 
{\frac{2g}{C}} =\mbox{const}.
\end{equation}
Thus, the particle on the brachistochrone moves in such a way that over time 
the rate of change in the inclination angle of its trajectory relative to 
the horizontal $x$-axis remains constant. The condition (\ref{AlphaTime60}) is satisfied 
both in the absence of sliding friction force and in the presence of the 
latter.

\end{document}